\documentclass[twocolumn, prl, aps, superscriptaddress,longbibliography,showpacs,amsmath,amssymb,floatfix]{revtex4-2}

\usepackage[utf8]{inputenc}   
\DeclareUnicodeCharacter{2009}{\,}

\usepackage{graphicx}
\usepackage{amssymb}
\usepackage{amsmath}
\usepackage{epsfig}
\usepackage{color}
\usepackage{tabu}
\usepackage{mathtools}
\usepackage[colorlinks,linkcolor=blue,anchorcolor=blue,citecolor=blue,urlcolor=blue]{hyperref}
\usepackage{physics}
\usepackage{float}
\usepackage{diagbox}

\begin{document}
\title{Coupling between vibration and Luttinger liquid in mechanical nanowires}
\author{Zeyu Rao}
\affiliation{Key Laboratory of Quantum Information, University of Science and Technology of China, Hefei, 230026, China}
\author{Yue-Xin Huang}
\email{hyx@gbu.edu.cn}
\affiliation{School of Sciences, Great Bay University, Dongguan 523000, China}
\author{Guang-Can Guo}
\affiliation{Key Laboratory of Quantum Information, University of Science and Technology of China, Hefei, 230026, China}
\affiliation{Synergetic Innovation Center of Quantum Information and Quantum Physics, University of Science and Technology of China, Hefei, Anhui 230026, China}
\affiliation{ Center for Excellence in Quantum Information and Quantum Physics}
\author{Ming Gong}
\email{gongm@ustc.edu.cn}
\affiliation{Key Laboratory of Quantum Information, University of Science and Technology of China, Hefei, 230026, China}
\affiliation{Synergetic Innovation Center of Quantum Information and Quantum Physics, University of Science and Technology of China, Hefei, Anhui 230026, China}
\affiliation{ Center for Excellence in Quantum Information and Quantum Physics}
\date{\today}

\begin{abstract}
	The vibration of the mechanical nanowire coupled to photons via photon pressure and coupled to charges via the capacity has been widely explored in experiments in the past decades. This system is electrically neutral, thus its coupling to the other degrees of freedom is always challenging. Here, we show that the vibration 
	can slightly change the nanowire length and the associated Fermi velocity, which leads to coupling between
	vibration and Luttinger liquid. We consider the transverse  and longitudinal vibrations of the nanowires, showing that the transverse vibration is much more 
	significant than the longitudinal vibration, which can be measured through the sizable frequency shift.  We predict an instability of the vibration induced 
	by this coupling when the frequency becomes negative at a critical temperature for the transverse vibrations in nanowires with low Fermi energy, which can be
	reached by tuning the chemical potential and magnetic field. The time-dependent oscillation of the conductance, which directly measures the Luttinger parameter, can provide evidence for this  coupling.  Our theory offers a new mechanism for exploring the coupling between the vibration and the electronic excitations, which may lead to intriguing applications in cooling and controlling the mechanical oscillators with currents. 
\end{abstract}
\maketitle

The quantum resonator can be fabricated in experiments with advanced fabrication techniques. It has been widely pursued in a cavity 
optomechanics \cite{aspelmeyer_cavity_2014, Bachtold2022Mesoscopic} and mechanical beams \cite{craighead_nanoelectromechanical_2000, Seung2007Synchronized,  ekinci_nanoelectromechanical_2005, Dinh2024Micromachined}, in which the parameters and the associated vibration frequencies can vary over several orders of magnitude (see Table I in Ref. \cite{Rosiek2024Quadrature}).
Applications of these systems, including ultrasensitive force 
sensing \cite{gavartin_hybrid_2012,li_ultra-sensitive_2007}, mass sensing \cite{lassagne_ultrasensitive_2008, chaste_nanomechanical_2012}, subfemtometre displacement 
sensing \cite{naik_cooling_2006}, and even charge and spin sensing \cite{RugarSingle2004},  have been demonstrated. In quantum metrology, it can be used as a tool for spin squeezing, 
entanglement, and even a useful platform for exploring the boundary between quantum and classical mechanics \cite{schwab2005putting}.
Therefore, the cooling of these systems down to the quantum limit is always the central issue at the forefront of these 
researches \cite{schliesser_resolved-sideband_2009,park_resolved-sideband_2009, oconnell_quantum_2010,zippilli_cooling_2009,urgell2020cooling,wen2020coherent,Bachtold2022Mesoscopic}.  Further, the strong nonlinearity in these systems can lead to bistability \cite{dorsel_optical_1983,purdy_tunable_2010},
synchronization, and even long-range coupling between different resonators \cite{luo_strong_2018, shim_synchronized_2007}.

The coupling of this vibration to the other physical degrees of freedom is essential for its application. In cavity optomechanics, the vibration is
coupled to the optical pressure \cite{marquardt_quantum_2007, kippenberg_cavity_2008}. There are also two different ways for the coupling
between vibration and electrons. The strain induced by the deformation may lead to charge-vibration coupling \cite{massel_microwave_2011,wollman_quantum_2015}
and spin-vibration coupling \cite{ovartchaiyapong_dynamic_2014,carter_spin-mechanical_2018,zippilli_cooling_2009,
whiteley_spinphonon_2019,Vigneau2022Ultrastrong,Huttner2023Optomechanical}. Via this mechanism, recently, the spin-vibration coupling has been demonstrated in quantum 
dots \cite{carter_spin-mechanical_2018,whiteley_spinphonon_2019,Vigneau2022Ultrastrong,Huttner2023Optomechanical}. The spring constant in the resonator may also be changed through the capacitive coupling,
based on which the single electron tunneling through the resonator can be directly reflected from the jump of vibration frequency \cite{teufel_circuit_2011,dobrindt_parametric_2008,Vigneau2022Ultrastrong,samanta2023nonlinear}.
Cooling of these resonators using current is still a great challenge  in experiments \cite{Stadlerground2014, zippilli_cooling_2009,urgell2020cooling,wen2020coherent}. 

\begin{figure}[h]
    \centering
    \includegraphics[width=0.49\textwidth]{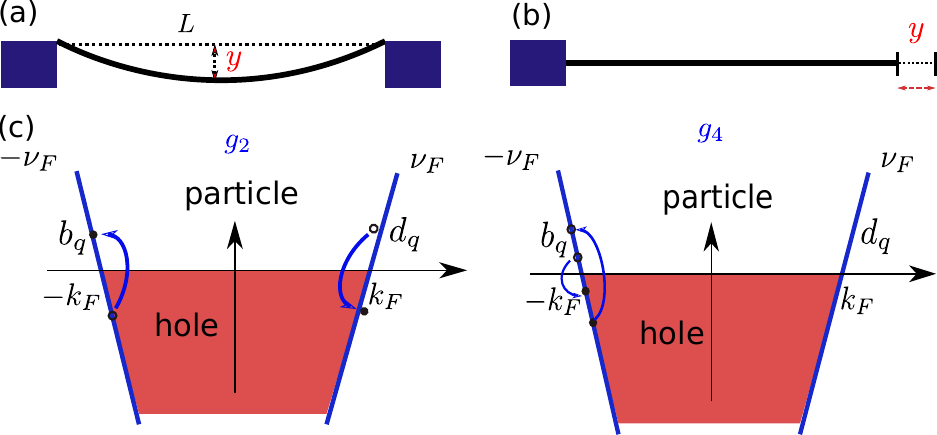}
	\caption{Physical model and the excitations in the Luttinger liquid. (a) and (b) correspond to the T-vibration and L-vibration 
	along the mechanical nanowire. The shaded black box corresponds to the clamped ends, and $y$ represents the displacement amplitude. 
	(c) Two major scattering channels near the Fermi points, where $g_2$ and $g_4$ are measured in units of $2\pi \nu_0$ (see Eq. \ref{eq-Hdiag}). }
    \label{fig-fig1}
\end{figure}

In this work, we aim to explore the direct coupling between the vibration and the plasmon excitation \cite{nemeth2022direct,wang2021gate} in the Luttinger liquid from the
vibration-induced change of nanowire length and the associated change of Fermi energy. This coupling induces a remarkable frequency 
shift of the mechanical nanowire, which exhibits some universal scaling laws at high temperature. We also predict the instability of
the vibration induced by this coupling at finite temperature, which can be measured in experiments even at low temperature by tuning the 
chemical potential and magnetic field. The possible parameters for these predictions are also examined using the available materials. This 
new coupling may lead to ‌intriguing‌ applications, such as cooling of the mechanical nanowires and the control of the vibration by current 
and the external electric and magnetic fields, which may also be generalized to other degrees of freedom, such as magnons \cite{zhang2016cavity,Shen2022Coherent,shen2025cavity,Hwang2024Strongly,Muller2024Temperature}.

We consider two typical    modes termed as transverse (T-) vibration and longitudinal (L-) vibration, as shown in Fig. \ref{fig-fig1}. 
The cantilever structure shares a similar feature with the bridge structure, which will be discussed briefly later.
The relevant interaction for the electron gases mainly takes place at the Fermi points $\pm k_F$ (Fig. \ref{fig-fig1}(c)). Let us 
introduce the subscripts L/R for the left and right movers, respectively, then the general Hamiltonian of the interacting spinless electron 
gases reads as \cite{miranda2003introduction, altland_condensed_2006,  giamarchi_quantum_2003,Fazzini2021Nonequilibrium}
\begin{eqnarray}
    H= \sum_k \left({\hbar^2 k^2 \over 2m^*} - \mu\right) c_{k}^\dagger c_{k}
    + \frac{\hbar}{2L}\sum_{ q}V_q \rho(q)\rho(-q), 
    \label{eq-interactingV}
\end{eqnarray}
where $\rho(q) = \sum_{k} c_{k+q}^\dagger c_{k}$, $m^*$ is the effective electron mass
and $\mu$ denotes the Fermi energy. When it is far away from half filling, the scatterings 
are mainly dominated by the forward scattering (denoted by $g_4$) and dispersion scattering (denoted by $g_2$) as shown in 
Fig. \ref{fig-fig1}(c). We can define the density excitations near the Fermi points as $\rho_r(q)=\sum_{k\sim \sigma_r k_F} c_{k+q}^\dagger c_{k}$,
where $\sigma_R=1$ and $\sigma_L=-1$ for $r  =R$,  $L$. 
These operators satisfy $[\rho_r(q), \rho^\dagger_{r^\prime}(q^\prime)] = -\delta_{rr^\prime}\delta_{q,q^\prime} \sigma_r \frac{qL}{2\pi}$. 
Let's define bosonic operators as $b_q^\prime=\sqrt{2\pi/qL}\rho_R(-q)$ and $d_q^\prime=\sqrt{2\pi/qL}\rho_L(q)$ for $q>0$, with $b_{-q}^\prime = b_q^{\prime\dagger}$ and 
$d_{-q}^\prime = d_q^{\prime\dagger}$, then via a Bogoliubov transformation we obtain \cite{miranda2003introduction, altland_condensed_2006,  giamarchi_quantum_2003,Fazzini2021Nonequilibrium}
\begin{eqnarray}
	H_\text{e}(L)=\hbar\sum_{q>0} \nu_\text{F} q \left( b_q^{\dagger}b_q+d_q^{\dagger}d_q\right),
    \label{eq-Hdiag}
\end{eqnarray}
where $\nu_\text{F} = \nu_0 \left[ (1+g_4)^2-g_2^2 \right]^{1/2}$, with $g_2, g_3 \ll 1$ for weak interaction.  Thus $\nu_0$ is the Fermi velocity without scattering. This mapping from boson and fermion 
particles is exact; thus, the interacting electron system can be described with a bosonic model \cite{giamarchi_quantum_2003,Rao2023Generalized,rao2025soliton,Entanglement2021Roosz,Density2022Roosz}.
This linear dispersion will not be affected by the disordered potential \cite{levy_luttinger-liquid_2006,gornyi_dephasing_2005}. 
Hereafter, these collective excitations will be termed as phonons \cite{zhu2022imaging,cohen2023universal,vianez2022observing,nemeth2022direct,wang2021gate,Hyun2025Band}. 

\begin{table}
    \centering
    \caption{Parameters for mechanical nanowires in experiments. $\Omega$ is in units of MHz, $E$ 
	is in unit of GPa, $L$, $w$ and $d$ are in unit of 
	$\mu$m, nm and nm, respectively, $\rho$ is in unit of g/cm$^3$, $T$ is in unit of 
	Kelvin used in experiments. The size of the nanotube is shown with its length and diameter.}
        \begin{tabular*}{1.0\linewidth}{@{\extracolsep{\fill}}ccccccc} \hline
         Ref. & $\Omega/2\pi$ & material & ($L, w, d$) & $E$ & $\rho$ & $T$ \\
        \hline
        \cite{massel_microwave_2011} & $32.5$ & Al & ($8.5,6,32$) &  69  &2.70 & $0.350$
        \\
        \cite{teufel_sideband_2011}& $10.6$ & Al & ($15,100$)  & 69 & 2.70 & $0.015$
        \\
        \cite{park_resolved-sideband_2009} & $123$ & Si & ($15,-,-$) &  165 & 2.32 & $10$
        \\
        \cite{verhagen_quantum-coherent_2012}& $78$ & $\,\mathrm{SiO_2}$ & ($15,-,-$) & 66 & 2.65 & 0.65
        \\
        \cite{thompson_strong_2008} & $0.134$ & SiN & ($1000,50,1000$) & 166 & 3.17  & 0.2
        \\
        \cite{lahaye2004approaching}& $19.7$ & SiN & ($8,200,20$)  & 166 & 3.17 & 0.056
        \\
        \cite{okamoto_coherent_2013} & $0.29$ & GaAs & ($100,400,-$) &  85.5 & 5.32 & 50
        \\
        \cite{etaki_motion_2008} & $2.0$ & InAs & ($50,42,4500$) & 51.4 & $5.68$ & $10$
		\\
        \cite{deng_strongly_2016} & $122$ & Nanotube & ($2,3,-$) &  1050 & 1.30 & 0.27
		\\
		\hline
	\end{tabular*}
    \label{tab-table1}
\end{table}

\begin{figure}[h]
    \centering
    \includegraphics[width=0.49\textwidth]{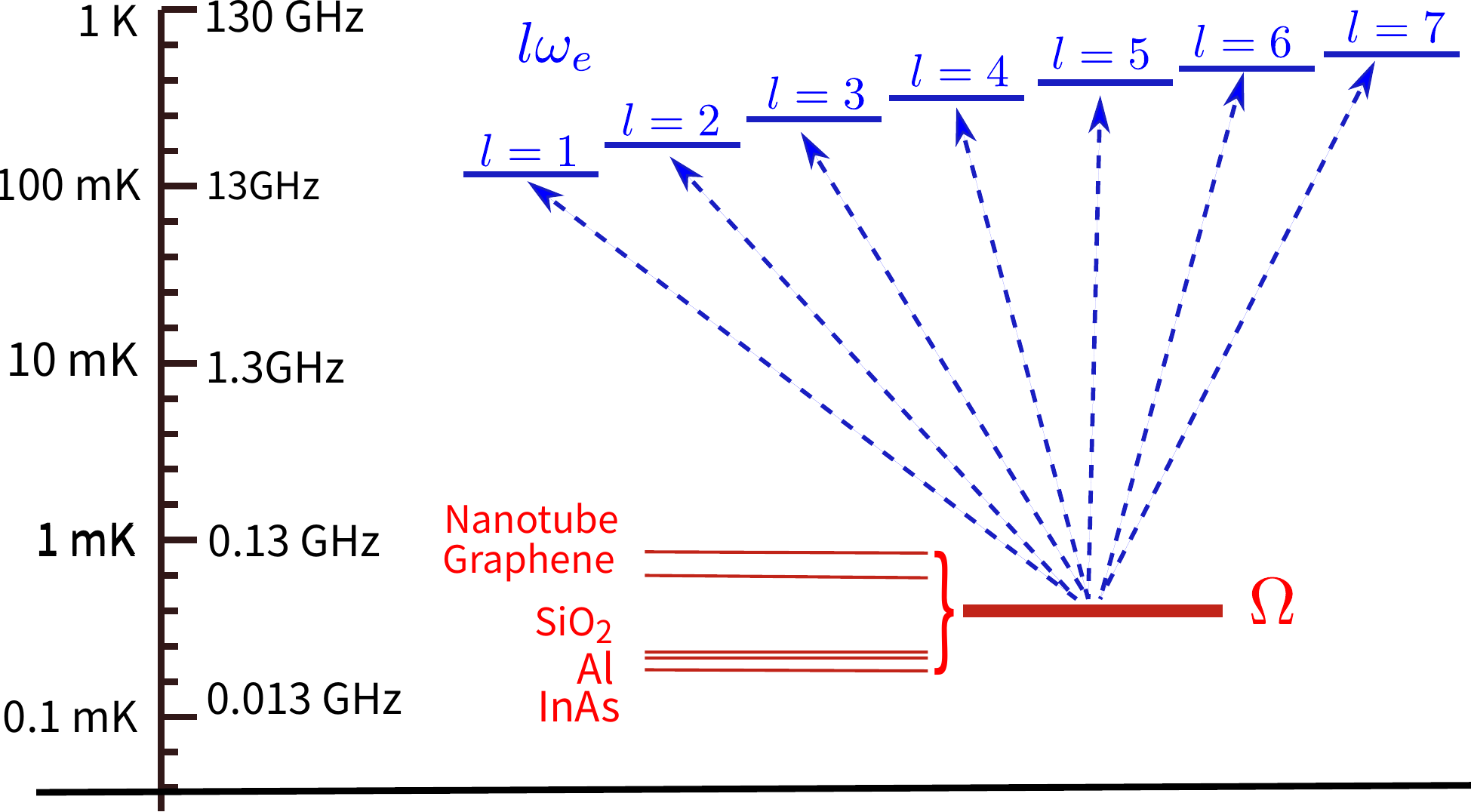}
    \caption{Energy scales of the vibration, plasmons, and temperature. 
	The typical materials are set in the same size $(3\,\mathrm{\mu m},10\,\mathrm{nm}, 15\,\mathrm{nm})$
	(notice that in nanotube, diameter $d = 4$ nm is used). The plasmon frequencies are denoted by $\omega_e$. 
	Here we set $\nu_\text{F} = 10^4$ m/s, thus $\omega_e\approx16$ GHz. This energy mismatch
	can not be decreased by increasing the system sizes.}
    \label{fig-fig2}
\end{figure}

The vibration can slightly change the length of the nanowire, while the total number of particles is fixed, thus 
$\nu_\text{F}$ should be length dependent. To determine the influence of this effect, we model the mechanical nanowire 
using the Euler-Bernoulli equation \cite{Purcell2002Tuning, Villanueva2013Nonlinearity}
\begin{equation}
	E I \partial_x^4 Y(x,t) =  -\rho A \partial_t^2 Y(x,t),
	\label{eq-euler}
\end{equation}
in which the frequencies for the T-vibration in Fig. \ref{fig-fig1}(a) can be expressed as $\Omega^\mathrm{(a)}_m= \beta_m^2L^{-2} \sqrt{E I/\rho A}$, 
 with $\beta_m \simeq (2m+1)\pi/2$, $I = A w^2/12$ is the geometric moment of inertia, $A = w d$ is the area of the cross section 
($w$ is the thickness and $d$ is the width), $\rho$ is the mass density, $E$ is the Young's module and $L$ is the static nanowire length. Note that 
the cantilever has similar frequencies, except some slightly different coefficients $\beta_m$, thus our 
prediction is also directly applicable to the physics of cantilevers \cite{cantileversbeta}. For the L-vibration in Fig. \ref{fig-fig1}(b), 
$\Omega^\mathrm{(b)}= \sqrt{E/\rho L^2}$. These parameters in several typical systems in experiments are summarized in Table \ref{tab-table1}.
Usually, $\Omega^\mathrm{(a)}/\Omega^\mathrm{(b)}= {\beta_m^2w}/{(\sqrt{12}L)} \ll 1$ when $L \gg w$, since bending needs much less energy than stretching for a rigid nanowire.

With the solution from Eq. \ref{eq-euler}, by assuming that the fluctuation magnitude is much smaller than the total 
nanowire length, we have 
\begin{equation}
    L^{\text{(a)}} = L + \alpha_m y^2/L, \quad     L^\mathrm{(b)} = L + y,
\end{equation}
where $\alpha_m \simeq m^2\pi^2/4$ for vibration $y$ along the transverse direction (see inset of Fig. \ref{fig-fig1}(a)) 
and along the longitudinal direction (see inset of Fig. \ref{fig-fig1}(b)). Let us introduce $y= y_\mathrm{\scriptscriptstyle ZP}
(a+a^\dagger)$ and $p=-i m_\mathrm{eff}\Omega y_\mathrm{\scriptscriptstyle ZP}(a-a^\dagger)$, where $a$ and $a^\dagger$ are bosonic operators,
and $y_\mathrm{\scriptscriptstyle ZP}=1/\sqrt{{2m_\mathrm{eff}\Omega}}$ is the zero point displacement with effective mass effective 
$m_\mathrm{eff} = \rho L A/3$.  The Hamiltonian of these two models can be written as $H_\text{R} = p^2/2m_\mathrm{eff}+1/2m_\mathrm{eff}\Omega^2 y^2$.
Therefore, we arrive at the total Hamiltonian for the coupling between the nanowire and the electrons as
\begin{equation}
	H = H_\text{R} + H_\text{e}(L + y).
	\label{eq-HRE}
\end{equation}

\begin{figure}[h]
    \centering
    \includegraphics[width=0.49\textwidth]{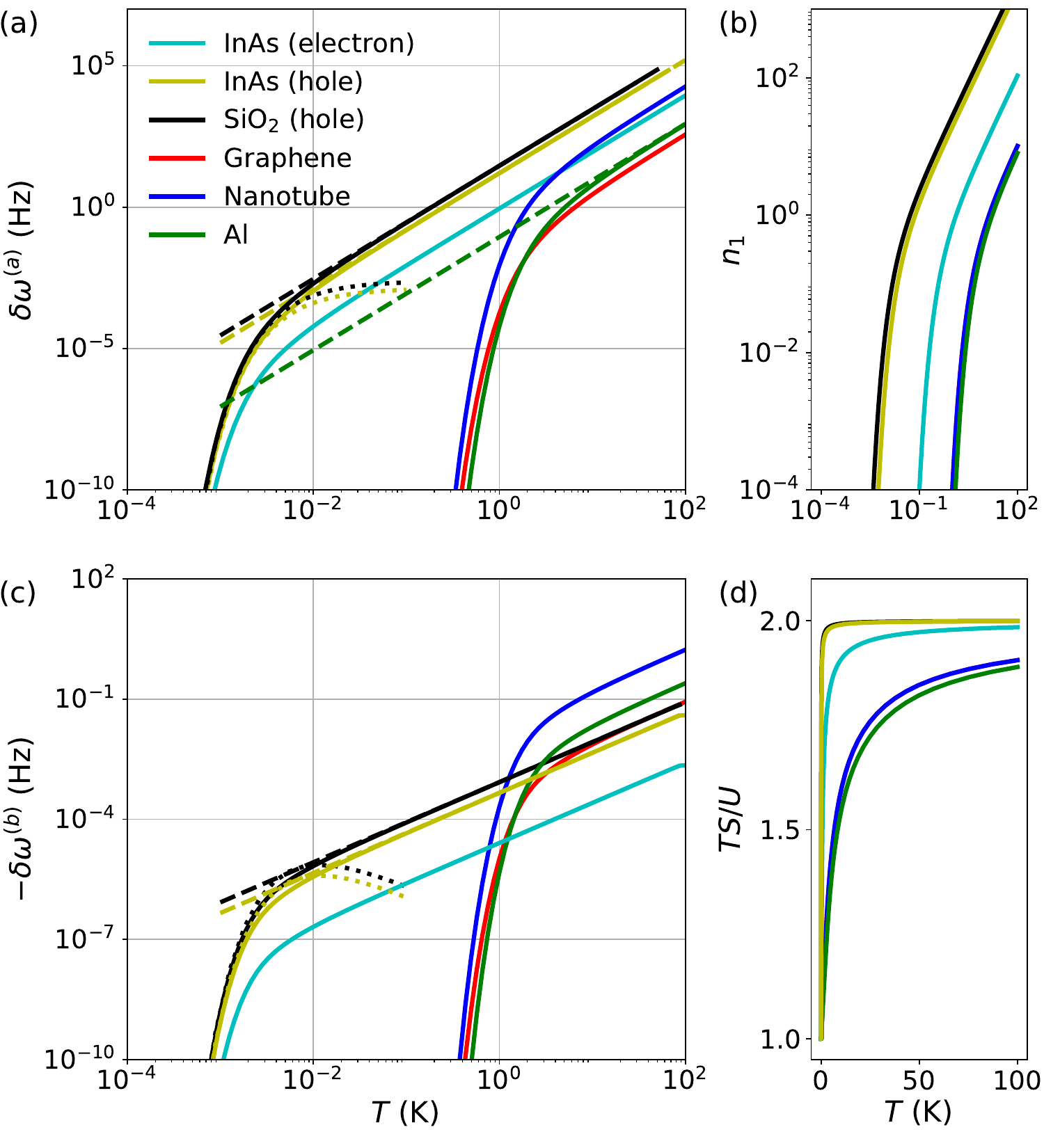}
	\caption{Frequency shifts in nanowires with various materials. In (a) and (c), the solid lines are numerical results based on 
        Eq. \ref{eq-heff}, and the dashed lines are our approximated frequency shifts using Eqs. \ref{eq-deltaomega1a} - \ref{eq-deltaomega2b} 
	for nanowires with the same sizes (100 $\mu$m, 10 nm, 15 nm). In this condition, $y_{\scriptscriptstyle ZP}$ is estimated to be of the order of 
	$10^{-12}$ m. In InAs, $m^* = 0.023m_\mathrm{e}$ and $0.41m_\mathrm{e}$ for electron and hole, respectively,
        with $n_c = 10^{15}$/cm$^3$ \cite{vurgaftman_band_2001}. The Fermi velocity 
	$\nu_\text{F}$ for InAs (electron), InAs (hole) and $\mathrm{SiO_2}$ (hole) are $7.9\times 10^4$, $4.4\times 10^3$, and 
	$3.1\times 10^3$ m/s, respectively; while in graphene, nanotube and Al, are $8\times 10^5$, $8\times 10^5$, 
	$1.0\times 10^6$ m/s. (c) Number of plasmons in the lowest mode. (d) $TS/U$, where $S$ is the entropy and $U$ is the 
	 Hamiltonian energy, as a function of $T$. When $T$ is high enough, this ratio will approach the upper bound 2 due to the 
	 linear nature of the collective excitations.}
    \label{fig-fig3}
\end{figure}

Eq. \ref{eq-HRE} is one of the major formulas we have derived in this work. Noticed that the change in length can slightly change 
the Fermi energy, which can influence $H_\text{R}$. This interaction, in most cases, is very weak; however, the number 
of plasmons is huge in the regime when $\Omega, \hbar \nu_\text{F} q \ll k_B T$, where $k_B$ is the Boltzmann constant 
(see Fig. \ref{fig-fig3}), which can lead to considerable influence on the vibration frequency. For small $y$, we have
\begin{eqnarray}
    H_\mathrm{e}&=&
    \hbar\sum_{q>0} \nu_\text{F}q \lambda \left( b_q^\dagger b_q+d_q^\dagger d_q \right),
    \label{eq-he}
\end{eqnarray}
where $\lambda=1-\alpha_m y^2/L^2$ for T-vibration and $\lambda=1-y/L+y^2/L^2$ for L-vibration.
Since $y \propto a + a^\dagger$, we see that these two different vibrations have totally different consequences for the nanowire.
In general, $\nu_\text{F} \pi/L \gg \Omega^{(a, b)}$ (see Fig. \ref{fig-fig3}), and this energy mismatch can not be decreased by 
increasing the system sizes. Thus, the resonance between the phonon mode and the vibration mode, analogous to the  parametric excitation of the pipeline by its conveying pulsating fluid, which can be observed in everyday life, is assumed not to be allowed in this manuscript. The possible resonance between them, via $\nu_\text{F} \pi/L = \Omega$, is estimated in Ref. \cite{resonantnote} and their related strong coupling will be discussed elsewhere. 

We mainly focus on the frequency shift at finite temperature induced by this coupling, which is observable in experiments. Typically, the coherent dynamics in the nanowire can persist with a typical 
time scale of the order of MHz, which can be much slower than the electronic relaxation time in the nanowire. In metals and semiconductors, the 
relaxation time near the Fermi surface can be of the order of picoseconds and sub-picoseconds \cite{liang_strong_2012,spisak_relaxation_2004}
as estimated from the free path length divided by $\nu_\text{F}$. 
For this reason, the interacting electrons can be treated as a background field in thermal equilibrium. 
From the partition function $Z_e = \text{Tr} e^{-\beta H_\text{e}} = e^{-\beta F}$, we can obtain the effective Hamiltonian for the nanowire
as  $H_\text{eff} = H_R + F = p^2/2m_\text{eff} + V_\text{eff}$, where 
\begin{eqnarray}
    V_\mathrm{eff}(y)&=& \frac{1}{2}m_\mathrm{eff}\Omega^2 y^2
    +\frac{2}{\beta}\sum_{l=1}^N \ln \left( 1-e^{-  l \eta\lambda} \right).
    \label{eq-heff}
\end{eqnarray}
Here $\eta=\omega_e/\omega_T$ is the ratio between the plasmon frequency $\omega_e= \nu_\text{F}\pi/L$ and the temperature frequency 
$\omega_T=k_BT/\hbar$. Notice
that here we can not directly set $b_q^\dagger b_q$ and $d_q^\dagger d_q$ to the mean number of plasmons because not only the mean energy but also the entropy
can contribute to $V_\text{eff}$. At high temperatures, the entropy from $F = U - TS$ is as important as the Hamiltonian energy $U = H$ (see Fig. \ref{fig-fig3}(d)).

The effective potential can qualitatively change the vibration frequency of the mechanical nanowire, providing a direct experimental signature for the coupling between the vibration and a Luttinger liquid. For $\eta\gg 1$, $e^{-l \eta\lambda}
\ll 1$, then by keeping only the leading term ($l = 1$) and for the two vibrations in Fig. \ref{fig-fig1}(a) and (b), we have
\begin{eqnarray}
    V_\mathrm{eff}^\mathrm{(a, b)}
    &\approx& \frac{1}{2}m_\mathrm{eff}\Omega^2 y^2 -\frac{2}{\beta}e^{-\eta}\eta (\frac{\alpha_m y^2}{L^2}, \frac{y}{L}- \frac{y^2}{L^2}).
    \label{eq-heff2}
\end{eqnarray}
At high temperature, $\eta\ll 1$ and $e^{-\eta\lambda}\approx 1$, it  is completely different. In this case, the plasmon occupation 
becomes important and we find
\begin{eqnarray}
    V_\mathrm{eff}^\mathrm{(a,b)}&=& 
    \frac{1}{2}m_\mathrm{eff}\Omega^2 y^2
    - \frac{\pi^2\hbar\omega_e}{3\eta^2}\left(\alpha_m\frac{y^2}{L^2}, \frac{y}{L}\right)
    \nonumber\\
    &&-\frac{2}{\beta}\left(\frac{\alpha_m y^2}{L^2},\frac{y}{2L}-\frac{y^2}{4L^2}\right).
    \label{eq-heff3}
\end{eqnarray}

We see that in these two limits, the coupling between the vibration and the Luttinger liquid gives rise to a new quadratic term, while the linear term is not important 
for the determination of modified frequency. Let us assume the new frequency as $\Omega'$, then we can define the frequency shift as 
$\delta \omega^{(a, b)} =  \Omega ^{(a, b)}-(\Omega')^{(a, b)}$. We are particularly interested in the condition when $\delta \omega$ is much smaller than 
$\Omega$. At low temperatures, we have
\begin{eqnarray}
    \delta\omega_1^\mathrm{(a)}&=& \frac{2\alpha_m}{\beta_m^2} e^{-\frac{\nu_\text{F} \pi\hbar}{k_BT L}} \frac{\hbar \nu_\text{F} \pi}{L^2}\sqrt{\frac{1}{\rho A E I}}, \label{eq-deltaomega1a} \\
    \delta\omega_1^\mathrm{(b)}&=& -e^{-\frac{\nu_\text{F} \pi \hbar}{k_B TL}} \frac{(\hbar \nu_\text{F} \pi)^2}{k_BTAL^4}\sqrt{\frac{1}{\rho E}}. \label{eq-deltaomega1b}
    \end{eqnarray}
And at high temperature
\begin{eqnarray}
    \delta\omega_2^\mathrm{(a)}&=& \frac{\alpha_m \hbar \pi}{3\beta_m^2\nu_\text{F}}\sqrt{\frac{1}{\rho A E I}}
    \left( \frac{k_BT}{\hbar} \right)^2, \label{eq-deltaomega2a} \\
    \delta\omega_2^\mathrm{(b)}&=& -\frac{k_BT}{2AL^2}\sqrt{\frac{1}{\rho E}}.
    \label{eq-deltaomega2b}
\end{eqnarray}
At low temperature, $\delta \omega \propto \exp(-\beta \nu_\text{F} \pi/L)$, thus the frequency shift is negligible. 
The increase in temperature can quickly increase the number of phonons, making the coupling 
more and more significant. At high temperature we find $\delta \omega_2 \propto T^\kappa$, with $\kappa = 2$ for T-vibration and 
$\kappa= 1$ for L-vibration. Especially, we find $\delta \omega_2^{(a)}$ is independent of nanowire length while $\delta \omega_2^{(b)}$ decreases with increasing of $L$.
These scaling laws can also be derived from the dimension analysis by assuming them to be a product of $\rho A$, $L$, $EI$, $\nu_\text{F}$, $k_B T$, 
$\hbar$ and total mass, which reflects the linear dispersion of the phonons, thus it is a general feature of the Luttinger liquids. 

With this theory, we present the frequency shifts in several typical materials in Fig. \ref{fig-fig3} using metals and nanotubes with $\nu_\text{F} \sim 10^6$ m/s
and semiconductors with tunable electron and hole concentration with $\nu_\text{F}$ from $10^3$ to $10^5$ m/s. In the normal metals, 
we have used concentration $n_\text{c} = 10^{22}\,\mathrm{/cm^3}$ and $m^* = m_e$ (rest electron mass); and for the semiconductor nanowires, 
$n_\text{c} \sim 10^{15} - 10^{18}$ $/\text{cm}^3$ \cite{vurgaftman_band_2001}. In general, the effective mass of hole is much heavier than the electron, thus 
has a much smaller $\nu_\text{F}$. We see that at low temperature, the frequency shifts are generally small,
thus are unquantifiable. The scaling laws of the frequency shifts at high temperature when $\omega_T \gg \omega_e$ can be measured in experiments. 
This simulation shows clearly that the larger $\nu_\text{F}$ is, the smaller frequency shift the system will be. Thus, in experiments, these kinds of shifts can be
more easily measured in semiconductor nanowires with a heavier effective mass. We also found that the T-vibration has a  much stronger frequency shift as compared with the L-vibration.

\begin{figure}
    \centering
    \includegraphics[width=0.49\textwidth]{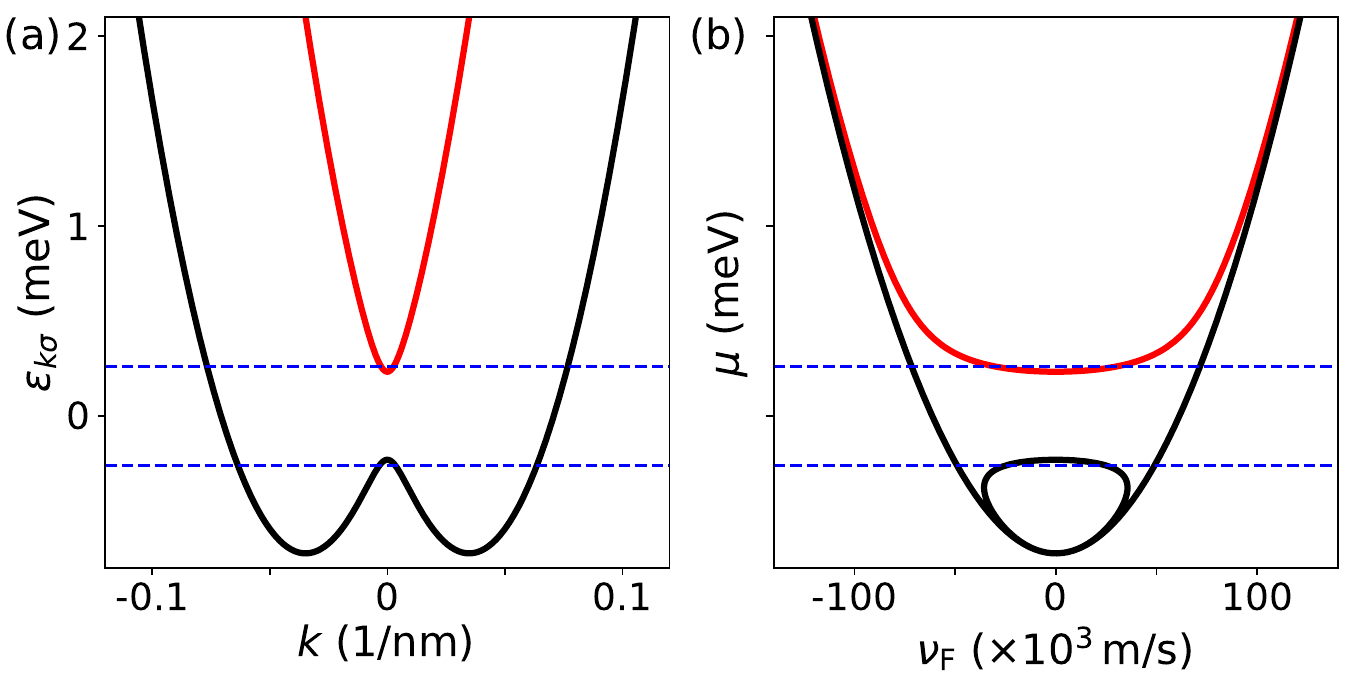}
    \caption{Tunability of $\nu_\text{F}$. (a) The band structure of InAs nanowire with spin-orbit coupling strength
    $\alpha\approx 40\,\mathrm{meV\cdot nm}$ \cite{liang_strong_2012, vurgaftman_band_2001} and magnetic field 
	$B_z=2\,\mathrm{T}$. (b) The corresponding $\nu_\text{F}$ at the Fermi points controlled by chemical potential. Near the 
	dashed lines, $\nu_\text{F}$ can be regarded as vanishing small.}
    \label{fig-fig4}
\end{figure}

A striking prediction in this coupling is the instability of the suspended bridge when the frequency becomes negative at
\begin{eqnarray}
    T>T_\text{c}= \frac{\beta_m^2}{k_BL}\sqrt{\frac{3\hbar \nu_\text{F} E I}{2\alpha_m \pi}}.
    \label{eq-Tomega}
\end{eqnarray}
It means, the longer the chain is, the smaller the temperature will be for this kind of instability. As naively expected, a nanowire 
with a larger Young's modulus and larger cross-section is expected to have a larger critical temperature. For 
the typical parameters in Fig. \ref{fig-fig3}(a), one can find $T_\text{c}$ is of the order of tens of Kelvin to hundreds of Kelvin \cite{crtical_temperature}.
This instability is more likely to be realized in samples with small $\nu_\text{F}$, long length $L$, and small Young's modulus. In realistic 
experiments, it is expected to happen at $T < T_\text{c}$, since the variance of the frequency shift can be calculated as 
\begin{eqnarray}
    \sigma(\delta \omega^\mathrm{(a, b)})= 3 \delta\omega^\mathrm{(a,b)},
	\label{eq-deltaomegaf}
\end{eqnarray}
where, again, the prefactor $3$ is exact only for linear dispersion. Thus, during the frequency 
shift, significant broadening of the spectra is also accompanied, which also scales as $T^{\kappa}$.
Thus, when the temperature is high enough, which is still smaller than $T_\text{c}$, we expect the instability of the nanowire about the 
zero point. In this analysis, we ignore the effect of gravitational force for the reason that it is a linear function of displacement $y$, 
which will not directly influence the instability of the nanowire. This instability can be useful for exploring the nonlinearity induced
bistability \cite{luo_strong_2018,purdy_tunable_2010, shim_synchronized_2007}.

This analysis also implies that the frequency shift can be measured and controlled by the current through the nanowire. Due to the fast carrier relaxation in
the nanowire, we can treat the vibration as an adiabatic process for the calculation of current and the related conductance. Let
us consider the T-vibration by adding a small bias over the two ends, thus the conductance is $G = K e^2/\hbar$ \cite{giamarchi_quantum_2003},
where $K = 2\pi \nu_\text{F} \sqrt{(1+ g_4 -g_2)/(1+ g_4  +g_2)}$ is the Luttinger parameter. This conductance may exhibit time-dependent
oscillation with period given by the vibration frequency, which is more easily distinguished in the nonlinear
regime with strong vibration \cite{carter_spin-mechanical_2018}. This method has been used in experiments to reveal the spin-vibration 
coupling \cite{carter_spin-mechanical_2018,whiteley_spinphonon_2019}.

Finally, we discuss the tunability of the coupling between vibration and the Luttinger liquid. From the frequency shift, we see that the 
only tunable parameter for a given sample is the Fermi velocity $\nu_\text{F}$, while all the other parameters are fundamental properties of the  nanowire,  thus can not be tuned. To solve this dilemma, let us generalize the spinless model to the 
spinful one by considering the physics in the InAs nanowire. The Hamiltonian can be written as $H = \hbar^2 k^2/(2m^*) + B_z \sigma_z + 
\alpha k \sigma_x$ \cite{miranda2003introduction}, where $\alpha$ is the spin-orbit coupling strength, and $B_z$ is the Zeeman splitting by the magnetic field.  The single particle spectra and the corresponding $\nu_\text{F}$ are shown in Fig. \ref{fig-fig4}. The gap width 
between the two bands can be controlled by the magnetic field. We find a significant change of $\nu_\text{F}$ from the order of $10^5$ m/s to 
$\nu_\text{F} \sim 0$ when $\mu$ approaches some van Hove points. In this regime the concentration density 
$n_\text{c} \sim 10^{18}$/cm$^3$. This tunability has been used in experiments for the searching of Majorana zero 
modes with fine tuning of chemical potential \cite{mourik_signatures_2012,Lutchyn2010Majorana, Sau2010Generic, Qiao2024Dressed}, thus our predicted instability can be observed in experiments by tuning the chemical potential and magnetic field with fixed temperature. 

To conclude, we present a new mechanism for coupling between the  vibration and the collective excitation of interacting electrons 
in the mechanical nanowire. This work can lead to numerous ramifications. For example, this theory can yield new frontiers for exploring intriguing couplings between vibration and electronic excitations, as
well as their cooling and control by current and external fields. It also opens the avenue for exploring the coupling between vibration and interacting Fermi liquids, magnons \cite{zhang2016cavity,Shen2022Coherent,shen2025cavity,Hwang2024Strongly,Muller2024Temperature}, and even with some phase transitions \cite{giamarchi_quantum_2003}, which certainly can greatly enrich the family of mechanical vibration physics \cite{aspelmeyer_cavity_2014, Bachtold2022Mesoscopic}. 

\textit{Acknowledgments}: This work is supported by the Strategic Priority Research Program of the Chinese Academy of Sciences (Grant No. XDB0500000), the Innovation Program for Quantum Science and Technology (2021ZD0301200, 2021ZD0301500), the National Natural Science Foundation of China (Grant No. 12474041), and the Pearl River Talent Recruitment Program (2023QN10X746). We thank Hao Yuan, Guang Wei Deng, Jie Qiao Liao, and Jian Hui Dai for valuable discussions.

\bibliography{ref}

\begin{thebibliography}{77}
\bibitem [{can()}]{cantileversbeta}%
  \BibitemOpen
  \href@noop {} {}\bibinfo {note} {The eigenfrequencies of the cantilever
  structure can be also written as $\Omega_m={\beta_m^2}\sqrt{{EI}/{(\rho
  A)}}/L^2$, where the coeffients $\beta_m$ are $1.8751$, $4.6941$, $7.8548$,
  $\cdots$. For $m\ge 3$, we can have an approximation expression as
  $\beta_m=(2m-1)\pi/2$.}\BibitemShut {Stop}%
\bibitem [{crt()}]{crtical_temperature}%
  \BibitemOpen
  \href@noop {} {}\bibinfo {note} {For a InAs nanowire with sizes of
  ($100\,\mathrm{\mu m},10\,\mathrm{nm},15\,\mathrm{nm}$), we have
  $\Omega_1^{(a)} \approx 0.19$ MHz. When $\nu_\text{F}=4.4\times 10^3$ m/s,
  the critical temperature is $T_\text{c}\approx 38.8$ K.}\BibitemShut {Stop}%
\end{thebibliography}
\end{document}